\documentclass[twocolumn]{revtex4}
\usepackage{graphicx,amssymb}
\usepackage{dcolumn}
\usepackage{bm}
\usepackage{longtable}
\usepackage{epsfig}

\def\prl{Phys. Rev. Lett. }
\def\prb{Phys. Rev. B }
\def\apl{Appl. Phys. Lett. }
\def\jap{J. Appl. Phys. }

\def\nim{Nucl. Instrum. Methods }
\def\nimb{Nucl. Instrum. Methods Phys. Res. B }

\begin{document}

\title{The mechanism of ion induced amorphization in Si}

\author{H. P. Lenka,$^1$ U. M. Bhatta,$^1$ P. K. Kuiri,
$^1$ G. Sahu,$^1$ B. Joseph,$^{1,}$\footnote{Present 
address:Istituto Tecnologie Avanzate, Contrada Milo 
91100, Trapani, Italy} B. Satpati,$^2$ and 
D. P. Mahapatra$^{1,}$\footnote{Electronic mail: 
dpm@iopb.res.in}}
\affiliation{$^1$Institute of Physics, Bhubaneswar 751005, India\\ 
$^2$ Institute of Minerals and Materials Technology, Bhubaneswar -751 013, 
India}  

\begin{abstract}

Some results on damage build up in, and amorphization of, Si, 
induced by 25-30 keV Al$_5^-$, Si$_5^-$ and  Cs$^-$ ions, at 
room temperature, are reported. We show that at low energy, 
amorphization is a nucleation and growth process, based on
the direct impact mechanism. With an Avrami exponent $\sim 1.6$, 
the growth towards amorphization seems to be diffusion limited. 
A transition to a completely amorphized state is indicated at 
a dose exceeding 17 eV/atom, which is higher than 6-12 eV/atom 
as predicted by simulations. The observed higher threshold 
could be due to temperature effects although an underestimation 
of keV-energy recoils, in simulation, may not be ruled out. 

\end{abstract}
\date{\today}
\maketitle

The study of ion implantation induced damage and recovery is a 
very important area of research in semiconductor processing, 
particularly involving Si. In all the cases involving doping 
through ion implantation, a damage layer is formed which 
must get back to a defect free crystalline state for any later 
application. In view of this, defect production by ions, 
together with its growth and annealing behavior, constitute 
an important area of study. 

Ion implantation induced amorphization has been the subject 
of intense research in which there is a long standing debate 
dating back to the seventies \cite{review,nordlund2}. One 
of the view points is that amorphization is caused from 
overlapping of amorphized pockets formed from defects created 
by individual ion cascades or {\it direct impact}. This is the 
so called {\it heterogeneous amorphization} as suggested by 
Morehead and Crowder \cite{morehead}. There are experimental 
data in support of this \cite{roult,howe,narayan}. Competing 
with this, there is a {\it homogeneous amorphization} mechanism 
where passage of the energetic ions results in the formation 
of a large number of {\it bond defects} (consisting of isolated 
point defects and interstitial-vacancy complexes). During 
implantation, with formation energies of $\sim$3 eV, these 
defects can be uniformly produced in the system and when 
their concentration increases beyond a certain limit the 
lattice becomes unstable resulting in a collapse to an 
amorphized state \cite{marques}. There are experimental data 
in support of this as well \cite{motooka,swanson,bai,holland} 
which also suggest amorphisation in Si is more like a phase 
transition \cite{hara2} induced by an accumulation of a 
sufficient number of defects. Reference \cite{review} provides 
a recent review on the subject.

In this letter we present some results regarding damage 
production and growth from low energy Al$_5^-$, Si$_5^-$ 
and a similar mass Cs$^-$ ions in Si at room temperature. 
Transmission electron microscopy (TEM) and channeling 
Rutherford backscattering spectrometry (RBS/C) have been
used for sample characterization. Increase in fluence, 
$\phi$, results in an increase in the number of atoms 
displaced from lattice positions leading to an increase 
in backscattered yield in the aligned condition. 
The growth in amorphous fraction, $f_a$, is obtained 
through a relative growth in the surface peak intensity, 
over and above that of a virgin sample. Three different 
growth models, all based on the direct impact mechanism, 
have been used to explain the data. We show, at low energy, 
amorphisation proceeds via a nucleation and growth process 
under the direct impact mechanism. The growth against dose 
is consistent with a Kolmogorov-Johnson-Mehl-Avrami (KJMA) 
equation with an exponent $\sim 1.6$ as against a value between 
3 and 4,predicted for 3D growth with a constant or slowly varying 
nucleation rate \cite{kjma}. This suggests the growth to be 
diffusion limited. Complete amorphization is found to 
occur above a dose, $D$, of 17 eV/atom, much higher 
than 6 - 12 eV/atom as suggested by Molecular Dynamics 
(MD) simulations \cite{nordlund2}. At doses below the 
saturation level, the amorphous to crystalline (a/c) 
interface is found to be rough. Since the experiments 
were carried out at room temperature, the rough 
interface can not be due to immobile defects as 
observed in case of a similar study \cite{elgohr} 
carried out at liquid N$_2$ temperature. However, at 
$\sim 50$ eV/atom a completely relaxed amorphous phase, 
with a smooth a/c interface, has been observed. 

The sample preparation involved both cluster (Al$_5$ 
and Si$_5$) and single atom (Cs) implantation into 
Si(100) (p-type, 20 $\Omega cm$) substrates, at room 
temperature, at low beam currents of $\sim$ 2-3 nA. 
Five samples were implanted with Si$_5^-$ ions, at 5 
keV/atom, to cluster fluence of $2\times 10^{11}$, 
$4\times 10^{11}$, $4\times 10^{12}$, $1.2\times 10^{13}$ 
and $1\times 10^{14}$ cm$^{-2}$ respectively. Two more 
samples were implanted with 25 keV Cs$^-$ ions and 30 
keV Al$_5^-$ ions (6 keV/atom) to fluence of 
$6\times 10^{13}$ cm$^{-2}$ and $1.6\times 10^{13}$ 
cm$^{-2}$ respectively. Excepting few selected ones, 
subjected to high resolution (HR) cross sectional TEM 
(XTEM) imaging (at 200 keV), all the samples were subjected 
to RBS/C measurements using 1.35 MeV He$^+$ ions at 
a scattering angle of 130$^o$. All the implantation and 
measurements were carried out using the facilities at the 
Institute of Physics (IOP), Bhubaneswar.
\begin{figure}[h]
\begin{center}
\includegraphics[width=6.8cm,height=6.4cm]{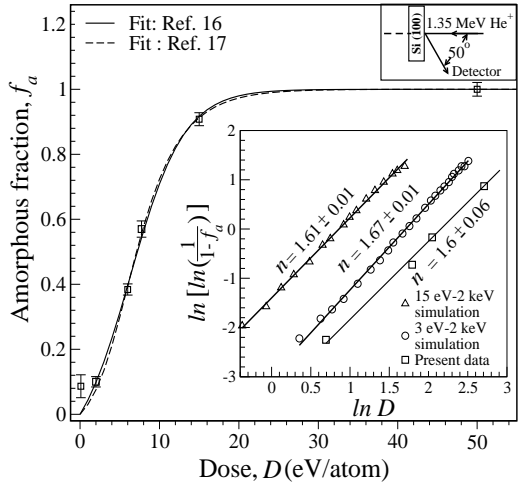}
\label{fig1}
\caption{Growth of amorphous fraction, $f_a$, with implantation dose,
$D$, in eV/atom. The fitted functions correspond to two of the models
as discussed in the text. KJMA fits to the present data and data
extracted from simulation results of Ref [2], are shown in the
inset. The three fits yield Avrami exponents, $n$, close to 1.6 as 
indicated. A schematic diagram of the RBS/C experimental geometry is
shown in the top right corner.}   
\end{center}
\end{figure}
For quantitative analysis, the amorphized fraction, $f_a$, 
for a given fluence $\phi$, was determined from the 
surface peak intensity in RBS/C data, relative to a 
virgin sample. This value was normalized to unity at 
the highest fluence (corresponding to complete 
amorphization). Since Al$_5$, Si$_5$ and Cs are 
different systems we convert the fluence, $\phi$ to 
dose, $D$ (eV/atom). This is done using the 
equation, $D=\phi E/(R_pN)$, $R_p$, $E$ and $N$ 
representing the projected range of the implanted atom, 
its energy and the atomic density of the matrix 
($5\times 10^{22}$ cm$^{-3}$) respectively. Since the clusters 
break almost immediately upon impact, the damage produced 
due to their implantation is mainly due to 5 or 6 keV Si 
or Al atoms respectively which have ranges of 10.2 and 
12.7 nm in Si \cite{trim}. Based on TEM data (shown later), 
for Si and Cs atoms, we take the ranges to be 10 nm and 20 nm respectively.
The results are shown in Fig. 1. The 95\% level in $f_a$ 
is seen to occur at a dose of $\sim 17.5$ eV/atom, complete 
saturation occurring around 20 eV/atom. This is higher than 
the saturation value of 6-12 eV/atom as indicated by simulations 
\cite{nordlund2}. 

The results can be explained in terms of a combined 
{\it Direct impact and defect stimulated mechanism} 
\cite{hecking}. This is mainly because presence of 
defects is known to result in a faster growth in the 
amorphized fraction \cite{bai,holland}. The cross-section 
for the defect stimulated process, $\sigma_s$, depends 
on the same for the direct impact process, $\sigma_a$. 
In such a case, 
$f_a=1-(\sigma_a+\sigma_s)/(\sigma_s+\sigma_a 
exp((\sigma_a+\sigma_s)D))$ \cite{weber}. A fit to the 
data yields values of 0.254 ($\pm$ 0.012) and 0.04 ($\pm$ 
0.005) for $\sigma_s$ and $\sigma_a$ respectively. 
\begin{figure}[h]
\begin{center}
\includegraphics[width=6.5cm,height=6.0cm]{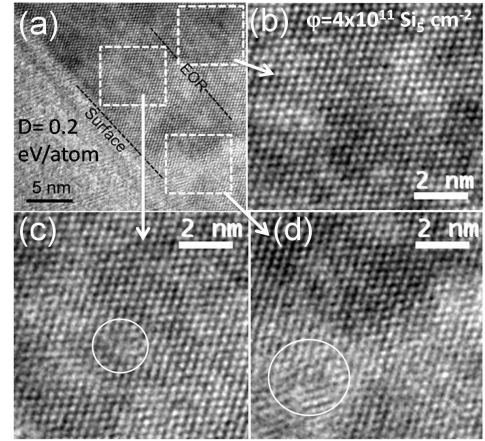}
\label{fig2}
\caption{(a) XTEM images of the Si sample irradiated with 
Si$_5$ to a cluster fluence, $\phi$, of $4\times 10^{11}$ 
cm$^{-2}$; (b) an HR image of the undamaged region as marked 
in (a); (c)\&(d) show two HR images of two marked boxed in
the defected region, showing (within circles) an amorphous 
patch and dislocations, within the ion EOR.}
\end{center}
\end{figure} 
\begin{figure}[h]
\begin{center}
\includegraphics[width=6.4cm,height=5.8cm]{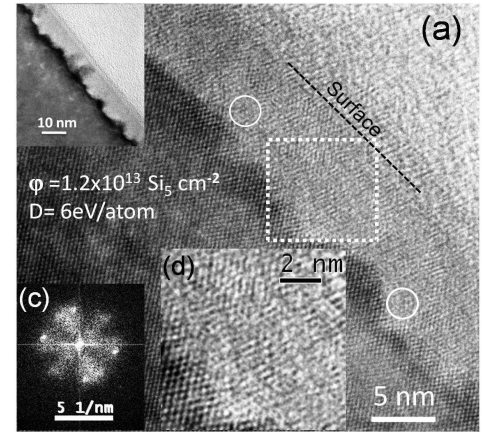}
\label{fig3}
\caption{(a) XTEM images of the Si sample irradiated with 
Si$_5$ clusters to a fluence, $\phi$, of $1.2\times 10^{13}$ 
cm$^{-2}$; (b) same as (a) at a lower resolution; (c) a 
Fourier transform of the region as marked by the box in (a); 
(d) the filtered image of the marked region. The circles 
indicate representative amorphous patches present in the 
damaged region.}
\end{center}
\end{figure} 
To look at cascade overlap effects we have used the 
{\it Cascade quenching and recrystallization} model 
proposed by Wang {\it et al} \cite{wang}. Here 
amorphization is assumed to be the result of direct 
impact of energetic ions, however, under a competing 
effect of partial recrystallization. Using a 
recrystallization efficiency, $A$, one can write 
$f_a = 1-1/\sqrt{A+(1-A)exp[2(1-A)D_n]}$, 
where $D_n=kD$, represents a dimensionless 
{\it normalized dose}. The parameter $k$ is related to 
the cross sectional area of the damaged region at any 
given dose. A fit using the above model yields a value 
of 0.893 ($\pm 0.002$) for $A$, together with a $k$ 
value of 2.214 ($\pm 0.048$). This $k$ value indicates, 
on an average, a 5 keV Si atom generates damage in a 
cylindrical region of cross sectional area of 22 nm$^2$ 
\cite{calc}, corresponding to a track radius of 
$\sim 2.65$ nm. Now we look at our fit values against 
the simulation results of Nord {\it et al} \cite{nordlund2} 
(as given in their table-IV). In our case saturation occurs 
at 20 eV/atom indicating a {\it normalized} dose, $D_n$ of 
about 44. With almost identical $A$ values, simulations 
for the two cases {\it viz} with 3 eV - 2.0 keV and 15 
eV - 2 keV recoils, show $k$ values of 3.822 and 6.326 
respectively. The corresponding $D$ values of 12 eV/atom 
and 6.5 eV/atom result in very similar $D_n$ values of 
45.86 and 41, in good agreement with present data. A $D_n$ 
value of 44 indicates a complete amorphisation at a 
$\phi$ of $4\times 10^{14}$ atoms-cm$^{-2}$ in good 
agreement with earlier data \cite{bai}. 
\begin{figure}[h]
\begin{center}
\includegraphics[width=6.4cm,height=5.8cm]{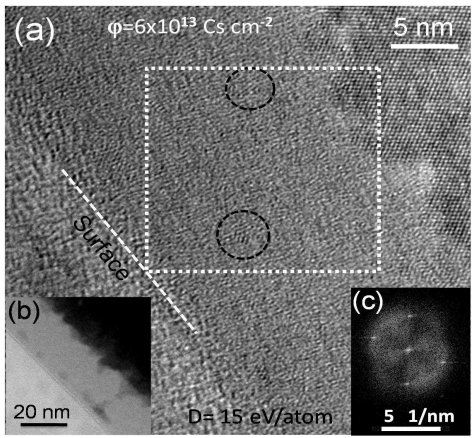}
\label{fig4}
\caption{(a) An XTEM image of the Si sample irradiated with 
Cs atoms to a fluence, $\phi$, of $6\times 10^{13}$ cm$^{-2}$; 
(b) same as (a) at a lower resolution; (c) a Fourier transform 
of the region as marked by the box in (a). The circles show 
some representative crystalline patches in a largely amorphized 
medium.}
\end{center}
\end{figure}
\begin{figure}[h]
\begin{center}
\includegraphics[width=6.4cm,height=5.8cm]{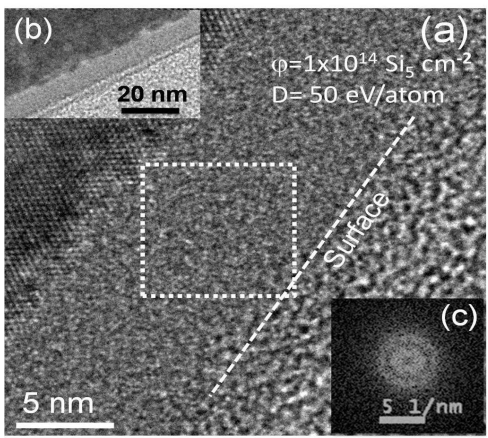}
\label{fig5}
\caption{(a) An XTEM image of the Si sample irradiated with 
Si$_{5}$ clusters to a fluence, $\phi$, of $1\times 10^{14}$ 
cm$^{-2}$; (b) same as (a) at a lower resolution; (c) a 
Fourier transform of the region as marked by the box in (a).}
\end{center}
\end{figure} 
The growth in $f_a$ with dose, $D$, can also be studied 
using a {\it Nucleation and growth} model based on the 
KJMA equation \cite{kjma} of the 
form $f_a = 1 - exp[-RG^{n-1}D^n]$. Here $R$ is related 
to the nucleation rate (dependent on $\sigma_a$) while 
$G$ is related to the growth velocity (dependent on
$\sigma_s$), at the interface \cite{weber}. The Avrami 
exponent, $n$, as determined from the slope of 
$ln[-ln(1-f_a)]$ {\it vs} $ln D$ (Fig. 1) yields a value 
1.6 $\pm 0.06$ (close to 3/2), indicating a diffusion 
limited nucleation and growth mechanism to be valid 
\cite{erukhi}. Simulation results of Nord {\it et al}, 
\cite{nordlund2}, for realistic (variable energy) recoil 
distributions, yield very similar $n$ values (Fig. 1) in 
agreement with our findings. This is in contrast to some 
earlier results on Ge induced amorphisation of Si carried 
out at higher energies, where $n$ was shown to be 3.5 
\cite{campisano}. Since $n$ has almost the same value in 
both simulated and experimental data, different $\sigma_s$ 
values indicates unequal growth velocities at the a/c 
interface. 

Fig. 2 (a) shows an XTEM picture of the defected region as 
obtained for an Si$_5$ fluence of $4\times 10^{11}$ cm$^{-2}$, 
corresponding to a dose, $D$, of 0.2 eV/atom. Fig. 2(b)
corresponds to an undamaged region. Fig. 2(c) \& (d) show
defects in the form of an amorphized patch and dislocations 
respectively, present within a depth of 10 nm (range of 5 keV 
Si in Si). HR XTEM images for the sample implanted with Si$_5$ 
fluence of $1.2\times 10^{13}$ cm$^{-2}$ (dose $\sim 6$ 
eV/atom) are shown in Fig. 3 (a) \& (b). In this case a heavily 
damaged region, extending from the ion end-of-range (EOR) up to 
the surface, can be seen. One can notice a rough interface 
together with a lot of crystalline patches in the near surface 
region. This is due to spatial overlap of collision cascades 
at the ion EOR where a high density of low energy recoils is 
expected to be produced. A Fourier transformed (FT) image of 
a 10 nm size region (Fig. 3 (c)) shows diffused 
$<111>$-diffraction spots along with shifted lobes indicating 
incomplete amorphisation and strain in the system. A Fourier 
inversion of the diffraction image (Fig. 3(d)) clearly shows 
the heavily damaged lattice with misoriented planes. One can 
also see some amorphized patches (shown in circles) in the 
damaged region. Compared to this, the XTEM images for the Cs 
implantation case (Fig. 4 (a) \& (b)), for a dose of $\sim 15$ 
eV/atom, show a greater degree of amorphization. The FT image 
(Fig. 4 (c)), of a damaged patch as marked, shows a continuous 
ring indicating a largely amorphized layer. The direct image
also shows this, however, with some crystalline patches still 
present. The corresponding fluence for 5 keV Si in Si, 
$1.5\times 10^{14}$ cm$^{-2}$, is very close to the threshold 
for amorphisation (the 95\% level occurs at 17.5 eV/atom). But 
the interface is still very rough. 
HR XTEM images of the damaged region obtained with an 
Si$_5$ implantation fluence of $1\times 10^{14}$ cm$^{-2}$ 
are shown in Fig. 5 (a)\&(b). This case corresponds to a 
dose of about 50 eV/atom which is well above the 
amorphization threshold. Fig. 5(a) shows complete 
amorphization, the diffraction image (Fig. 5(c)) of a 
representative marked region showing two clear rings. The 
a/c interface is also seen to be rather flat, mainly due 
to stress relaxation \cite{volkert}.

The above observations can be explained as follows. Si 
(or Al) atoms at 5 (or 6 keV) or 25 keV Cs atoms, in 
addition to creating point defects, produce amorphous 
patches in the implanted region. Some of these, are seen 
in Fig. 2(c). With increase in dose, with larger overlap of 
collision cascade regions there is a growth in the amorphized 
volume extending from the EOR towards the surface. Fig. 3 and 
Fig. 4 show this, with chunks of crystalline material present 
near the surface. These crystalline regions finally get amorphized 
from impact of low energy recoils due to successive ion passes. 
However, the growth in amorphisation is diffusion 
limited as the Avrami exponent, $n$, has a value close 
to 3/2 \cite{kjma}. We believe, the above diffusion 
limited growth mechanism also holds for amorphisation 
near the EOR. In this case the rough interface is 
smoothed out because of the depth limit imposed by the 
ion energy in addition to stress relaxation. 

For most of the potentials used for simulation of the 
Si structure, an energy deposition of 6-8 eV/atom is 
needed for getting an amorphous structure \cite{nordlund2}. 
This is small compared to the amorphizing dose ($>17$ 
eV/atom), as obtained here. One of the reasons could be 
that the experiments were carried out at room temperature 
($300 K$) where small amorphous patches can get 
recrystallized even without any ion passage \cite{delaRubia},
leading to a smaller growth rate. Another reason could be a 
possible underestimation of the recrystallization effects 
due to keV energy recoils in the simulation. At higher beam 
energies and higher beam currents, amorphisation will be suppressed by 
recrystallization induced by relatively larger number 
of recoils, pushing the amorphization threshold to a 
higher dose. This could be the reason behind getting an 
amorphization threshold of $4\times 10^{14}$ cm$^{-2}$, 
with 230 keV Si \cite{bai}, as compared to almost half 
that in the present case. Beam energy and current induced 
annealing effects may even result in recrystallization of 
amorphous patches leading to the formation of point defects 
and related complexes \cite{delaRubia,santos}. In case the 
concentration such defects becomes too high homogeneous 
amorphization may take place. This does not happen in
the present case.

To conclude, using 30 keV Al$_5$, 25 keV Si$_5^-$ and 
25 keV Cs$^-$ ions, we have shown that ion induced 
amorphisation in Si, at low energy, is very much dependent 
on the direct impact mechanism resulting in heterogeneous 
amorphisation. At these energies, complete amorphisation 
seems to take place through a diffusion limited, nucleation 
and growth process. 

We thank Prof. S. D. Mahanti, Michigan State University, East 
Lansing, USA and Prof. S. M. Bhattacharjee, IOP, Bhubaneswar 
for some critical comments and suggestions. We also thank the 
staff of the Ion Beam Laboratory, IOP, Bhubaneswar for their
help during the experiments.


\begin{thebibliography}{10}
\bibitem{review}L. Pelaz, L. A. Marques and J. Barbolla, \jap {\bf 96}, 5947 
	(2004) and references therein.
\bibitem{nordlund2} J. Nord, K. Nordlund and J. Keinonen, \prb {\bf 65},
	165329 (2002) and references therein.     
\bibitem{morehead}F. F. Morehead, Jr. and B. L. Crowder, Radiat. Eff. {\bf 6}, 
	27 (1970).
\bibitem{roult}M. O. Roult, J. Choumont, J. M. Penisson and
	A. Bourret, Philos. Mag. A {\bf 50}, 667 (1984).
\bibitem{howe} L. M. Howe and M. H. Rainville, \nim {\bf 182/183}, 143 (1981)
\bibitem{narayan} J. Narayan, D. Fathy, O. S. Oen and O. W. Holland, J. Vac. 
	Sci. Technol. A {\bf 2}, 1303 (1984), J. Narayan, O. S. Oen, D. Fathy 
	and O. W. Holland, Mater. Lett. {\bf 3}, 67 (1985)   
\bibitem{marques}L. A. Marques, L. Pelaz, J. Hernandez and J. Barbolla, 
        \prb {\bf 64}, 045214(2001); L. A. Marques, L. Pelaz,
        P. Lopez, I. Santos and M. Aboy, \prb {\bf 76}, 153201 (2007).
\bibitem{motooka}T. Motooka, S. Harada and M. Ishimaru, \prl 
        {\bf 78}, 2980 (1997).
\bibitem{swanson} M. L. Swanson, J. R. Parsons and C. W. Hoelke, Radiat. Eff. 
{\bf 9} , 249 (1971). 
\bibitem{bai} G. Bai and M. -A. Nicolet, \jap {\bf 70}, 648 (1991).
\bibitem{holland} O. W. Holland and S. J. Pennycook, \apl 
	{\bf 55}, 2503 (1989).
\bibitem{hara2}H. P. Lenka, B. Joseph, P. K. Kuiri, G. Sahu, P. Mishra, D. 
	Ghosh and D. P. Mahapatra, J. Phys. D: Appl. Phys. {\bf 41}, 315305 
	(2008).
\bibitem{kjma}A. N. Kolmogorov, Bull. Acad. Sci. USSR {\bf 3}, 355 (1937);
	English translation: Selected Works of A. N. Kolmogorov,
	edited by A. N. Shiryayev Kluwer, Dordrecht, 1992, Vol. {\bf 2},
	p. 188, W. A. Johnson and R. F. Mehl,
	Trans. Am. Inst. Min. (Metall.)Eng. {\bf 135}, 416 (1939); 
	M. Avrami, J. Chem. Phys. {\bf 7}, 1103 (1939).
\bibitem{elgohr}M. K. El-Gohr, O. W. Holland, C. W. White and S. J. Pennycook,
	J. Mater. Res. {\bf 5}, 352 (1990).
\bibitem {trim}J. F. Ziegler, J. P. Biersack and U. Littmark, 
        {\it The Stopping and Range of Ions in Matter} (Perganon Press, New 
        York, 1995).
\bibitem{hecking} N. Hecking, K. F. Heidemann and E. T. Kaat, \nimb 
	{\bf 15}, 760 (1986); 
\bibitem{wang} S. X. Wang, L. M. Wang and R. C. Ewing, \prb {\bf 63}, 
	024105 (2001).
\bibitem{weber}W. J. Weber, \nimb {\bf 166/167}, 98 (2000). 
\bibitem{calc} For any $D$, deviding $D_n$ by $\phi$ one can determine 
        the cross sectional area of the ion track. $\phi$ needs to be
        calculated taking the appropriate $E$ and $R_p$.
\bibitem{erukhi}V. Erukhimovitch and J. Baram, \prb {\bf 51}, 6221 (1995).
\bibitem{campisano}S. U. Campisano, S. Coffa, V. Raineri, P. Priolo and E. Rimini,
	\nimb {\bf 80/81}, 514 (1993).
\bibitem{lewis-niminen} L. J. Lewis and R. M. Nieminen, \prb {\bf 54}, 
	1459 (1996).
\bibitem{volkert}C. A. Volkert, \jap {\bf 70}, 3521 (1991).
\bibitem{santos}I. Santos, L. -A. Marques and L. Pelaz, \prb {\bf 74}, 
	174115 (2006).
\bibitem{delaRubia} T Diaz dela Rubia and G. H. Gilmer, \prl 
	{\bf 74}, 2507 (1995).
\end{thebibliography}
\end{document}